# Authentication and Authorization in Server Systems for Bio-Informatics

**K. Lakshmi Madhuri[1], T.R. Gopalakrishnan Nair [2]**
Research & Industry Incubation Center,
Dayananda Sagar Institutions, Bangalore 560 078 Karnataka, INDIA

**Abstract**
Authentication and authorization are two tightly coupled and interrelated concepts which are used to keep transactions secure and help in protecting confidential information. This paper proposes to evaluate the current techniques used for authentication and authorization also compares them with the best practices and universally accepted authentication and authorization methods. Authentication verifies user identity and provides reusable credentials while authorization services stores information about user access levels. These mechanisms by which a system checks what level of access a particular authenticated user should have to view secure resources is controlled by the system. Digital signatures and certificate can be used to authenticate the source of messages with the help of tls and ssl which are now an integral part of most web browsers clients and web servers. Digital signature and certificate is one of the frontier approaches which security experts recommend.Bio informatics servers are internet services for sequence analysis and the prediction of aspects of protein structure and function. Users submit protein sequence or alignments the server returns a multiple sequence alignment prosite sequence motifs low complexity regions seg prodom domain assignments nuclear localization signals regions lacking regular structure and predictions of secondary structure solvent accessibility globular regions transmembrane helices coiled coil regions structural switch regions and disulfide bonds. For all services users can submit their query either by electronic mail or interactively from world wide web. Since there would be highly confidential and secure data in certain type of disease research and drug discovery we need an effective authentication and authorization technique to safeguard the interest of the research.

## Introduction

The advancement of data processing and emergence of newer applications were possible, partially because of the growth of the robust servers. According to the Moore's law the number of transistors in a single microchip is double every 18 months, and the growth of the semiconductor industry has so far followed the prediction [6].

With most of this data travelling over shared IP networks, and regulations pushing to secure vulnerable traffic, security technologies must now be enhanced to provide an appropriate level of safety. In the near future, all network traffic, whether inside the corporate LAN, across the WAN, or over the Internet, won't be trusted. This means that all frames and packets must have their payloads inspected for malicious code and all traffic must be encrypted. Current encryption solutions simply do not scale to support the global problem of applying data protection at all end points. There are a number of encryption solutions deployed today to solve portions of the problems. There are, for example, application level encryption tools, SSL VPNs, IPSec VPNs, Layer 2 encryption (IEEE 802.1ae), file transfer encryption tools, telnet encryption, e-mail encryption tools--the list goes on. **Transport Layer Security** (TLS)/Secure Sockets Layer (SSL) is implemented between the application layer and the transport layer. Using TCP for reliable delivery, TLS/SSL primarily secures Web-based applications, although any TCP application can be secured [1].

Bioinformatics is a domain which is advancing in leaps and bounds. The bioinformatics server provides a universal portal to end users for bioinformatics related computing & data accessing and also supports some special bioinformatics applications. Since this science processes these data's and discover new biological principles by using computers [2]. The ability of a traditional supercomputer is not powerful enough to effectively manage, analyze and utilize such a lot of data. In this server many Bioinformatics applications are supported like AAGS, BLAST, TMHMM etc, thus it becomes more important to control the accesses to local or remote resources in a secure manner [3]. Before submitting a task the researchers must obtain a legal account from local administrator. Then they can execute tasks by calling resources authorized by the server. The server will refuse any request from one whose identification can not be verified or is not allowed to utilize it. Authenticating and Authorization are central concepts of any secured server. Authentication and Authorization are the mechanisms by which a system checks what level of access a particular authenticated user should have to secure resources controlled by the system.

**Overview of Bioinformatics application Server**

Bioinformatics web application server should consist these basic components some of the important components are listed below

- A source of data.
- An application programming language to access and analyze the data.
- A web application platform to provide a HTML user interface for your data and analysis results.
- Optionally, a data store, such as a relational database, to store results or user's data.
- Optionally, you would reuse software tools and libraries developed by others.

There are a number of bioinformatics databases that are publicly available and, in addition, make the databases and software tools themselves available for building your own web user interfaces. The following table summarizes the programming languages, web application platforms, and bioinformatics tools and libraries which can be used in the development of the server [7].

**Table 1: programming languages, web application platforms, and bioinformatics tools and libraries**

| Programming Language | Web Technology | Advantages | Disadvantages |
|---|---|---|---|
| Perl | Common Gateway Interface (CGI) | Large active open source community, cheap in an ISP hosted environment | Language not as structured as strongly typed languages |
| Java | Servlet / Java 2 Enterprise Edition | Large active open source community, well developed web programming model | Expensive in an ISP hosted environment |
| C and C++ | (CGI) | Some critical programs, such as BLAST, are already written using it, can be integrated with Perl | Not suitable for web application development (too complex) |
| C# | ASP.NET | Relatively cheap in an ISP hosted environment, Well developed web programming mode | Small active open source community, tools are not open source or free |
| PHP | Apache Module and others | Cheap in an ISP hosted environment | Smaller active open source community, language not as suitable to large scale software development |
| JavaScript (AJAX) | All other server side languages | Better user experience | Smaller active open source community, adds an additional level of software development complexity |

Web servers are actually better called HTTP servers because it receives and responds to Hypertext Transfer Protocol. Developing a web application you will be most concerned how your application programming language can interface with the web server. However, there are a number of aspects that are common to many application programming languages and some that are particular to certain web servers. Some of these aspects are file permissions and access, authentication and authorization [7]. Apache HTTP Server is reportedly the most commonly used web server. It is a favorite of Internet Service Providers (ISP's) and of the open source community. Apache has distinct ways of dealing with the question of whether a particular request for a resource will result in that resource actually is returned. These criteria are called *Authorization* and *Authentication*. Because these three techniques are so closely related in most real applications, it is difficult to talk about them separate from one another. In particular, authentication and authorization are, in most actual implementations, inextricable. Basic authentication is the simplest method of authentication and for a long time was the most common authentication method used. However, other methods of authentication have recently passed basic in common usage, due to usability issues that will be discussed in a minute. There are two configuration steps which you must complete in order to protect a resource using basic authentication or three, depending on what you are trying to do. Addressing one of the security caveats of basic authentication, digest authentication provides an alternate method for protecting your web content. However, it to has a few caveats.
The steps for configuring your server for digest authentication are very similar for those for basic authentication.

1. Create a password file
2. Set the configuration to use this password file
3. Optionally, create a group file

digest authentication has this great advantage that you don't send your password across the network in the clear, it is not supported by all major browsers in use today, and so you should not use it on a web site on which you cannot control the browsers that people will be using, such as on your intranet site.

There are also challenges which the web server may face because of the way that Basic authentication is specified, your username and password must be verified every time you request a document from the server. This is even if you're reloading the same page, and for every image on the page (if they come from a protected directory). As you can imagine, this slows things down a little. The amount that it slows things down is proportional to the size of the password file, because it has to open up that file, and go down the list of users until it gets to your name. And it has to do this every time a page is loaded.

A consequence of this is that there's a practical limit to how many users you can put in one password file. This limit will vary depending on the performance of your particular server machine, but you can expect to see slowdowns once you get above a few hundred entries, and may wish to consider a different authentication method at that time.

**Implementation of Authentication and Authorization in Bioinformatics Server**

In implementation situations, however, you will not know the IP addresses of all your clients. In this case, one straightforward approach to implementing authentication is to leverage the authentication features of the protocol used to exchange messages [8]. For most Web Services, this means leveraging the authentication features available for HTTP. Microsoft Internet Information Server 5.0 (IIS) supports several authentication mechanisms for HTTP shown in the following table:

**Table 2: Authentication Mechanisms for HTTP**

| | |
|---|---|
| Basic | Use for non-secure or semi-secure identification of clients, as username and password are sent as base64-encoded text, which |
| | is easily decoded. IIS will authorize access to the Web Service if the credentials match a valid user account. |
| Basic over SSL | Same as Basic authentication except that the communication channel is encrypted, thus protecting the username and password. A good option for Internet scenarios, however using SSL has a significant impact on performance. |
| Digest | Uses hashing to transmit client credentials in a secure way. However, may not be widely supported by developer tools for building Web Service clients. IIS will authorize access to the Web Service if the credentials match a valid user account. |
| Integrated Windows authentication | Useful primarily in Intranet scenarios. Uses NTLM or Kerberos. Client must belong to the same domain as the server, or belong to a trusted domain of the server's domain. IIS will authorize access to the Web Service if the credentials match a valid user account. |
| Client certificates over SSL | Requires each client to obtain a certificate. Certificates are mapped to user accounts, which are used by IIS for authorizing access to the Web Service. A viable option for Internet scenarios, although use of digital certificates is not widespread at this time. May not be widely supported by developer tools for building Web Service clients. Only available over SSL connections, thus performance may be a concern. |

In one aspect, the invention features a method for authenticating and authorizing a user to log onto to a network element includes providing user identifier and user authentication information to a centralized authority. The centralized authority is responsible for authenticating users attempting to log onto the network element. The method also includes the centralized authority authenticating the user based on the user identifier and the user authentication information. Additionally the method includes generating a response that includes the user identifier and a privilege level for the user.

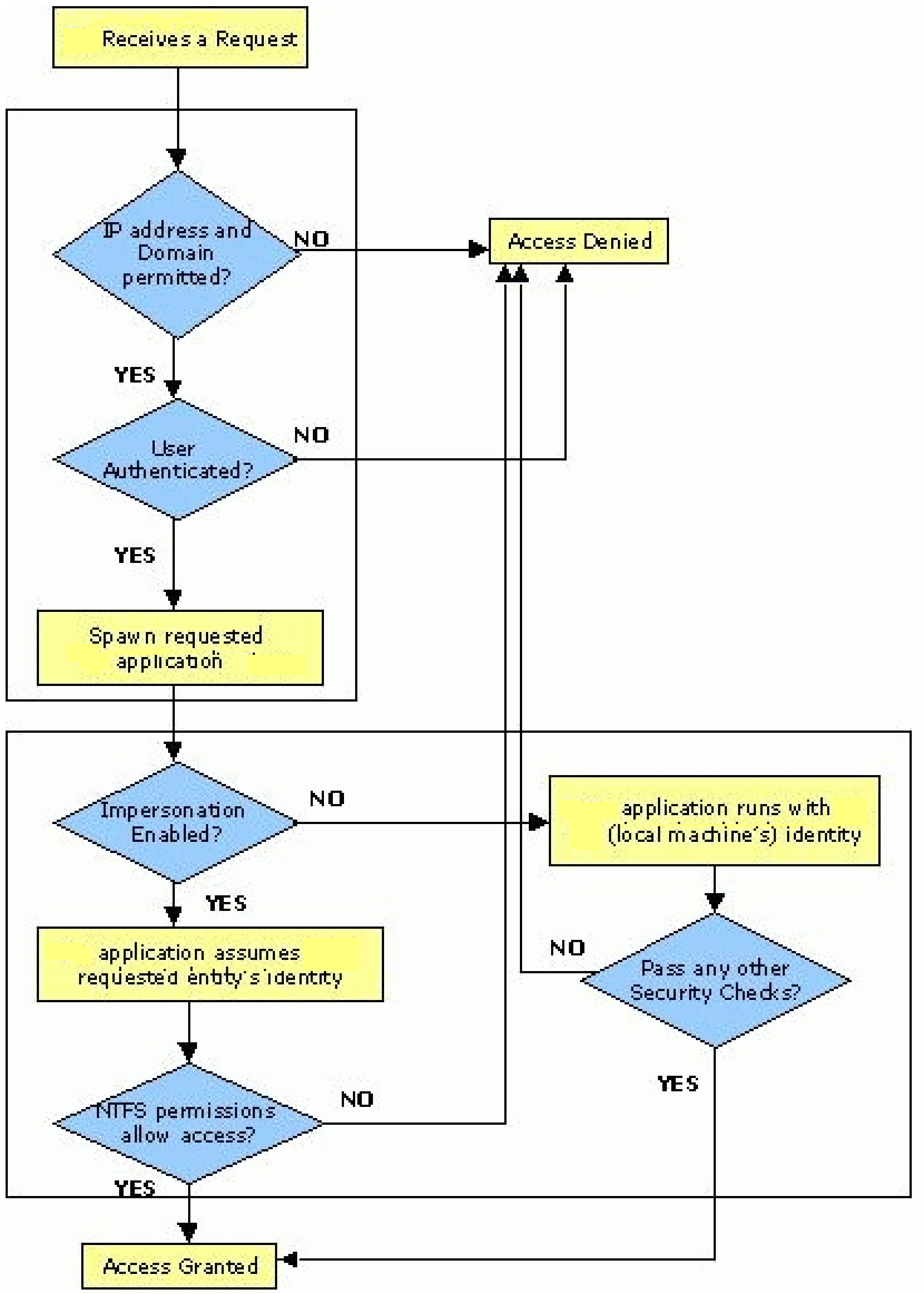

**Flow Diagram 1: Authentication & Authorization in Bioinformatics Server**

Authentication and Authorization in Web Server
There are three types of modules involved in the authentication and authorization process. You will usually need to choose at least one module from each group.

a) Authentication type
b) Authentication provider
c) Authorization

The Authorization module generally contains both authentication and authorization. The module Authentication provider is not an authentication provider by itself, but allows other authentication providers to be configured in a flexible manner. Network administrators frequently use the Lightweight Directory Access Protocol (LDAP) to implement a centralized directory server. You can use LDAP to authenticate users in Apache. The module mod_authnz_ldap in Apache server is both an authentication and authorization provider.

In the apache HTTP Server there will be 2 directives AuthType directive and Require directive in AuthType directive modules of Authentication type and Authentication Provider will be included.

- Authentication type (see the `AuthType` directive)
  - `mod_auth_basic`
  - `mod_auth_digest`
- Authentication provider
  - `mod_authn_alias`
  - `mod_authn_anon`
  - `mod_authn_dbd`
  - `mod_authn_dbm`
  - `mod_authn_default`
  - `mod_authn_file`
  - `mod_authnz_ldap`
- Authorization (see the `Require` directive)
  - `mod_authnz_ldap`
  - `mod_authz_dbm`
  - `mod_authz_default`
  - `mod_authz_groupfile`
  - `mod_authz_owner`
  - `mod_authz_user`

## Authentication and Authorization in Application Server

The Application Server supports four types of authentication. An application specifies the type of authentication it uses within its deployment descriptors JACC (Java Authorization Contract for Containers) is part of the Java EE specification that defines an interface for pluggable authorization providers. This enables the administrator to set up third-party plug-in modules to perform authorization [9].

**Table 3 : Application Server Authentication Methods**

| Authentication Method | Communication Protocol | Description | User Credential Encryption |
|---|---|---|---|
| BASIC | HTTP (SSL optional) | Uses the server's built-in pop-up login dialog box. | None, unless using SSL. |
| FORM | HTTP (SSL optional) | Application provides its own custom login and error pages. | None, unless using SSL. |
| CLIENT-CERT | HTTPS (HTTP over SSL) | Server authenticates the client using a public key certificate. | SSL |

OpenSSO Enterprise integrates core features such as access control through authentication and authorization processes, and federation. These functions can be configured using the administration console or the `ssoadm` command line utility. Figure 2 is a high-level illustration of the interactions that occur between parties (including the policy agent, browser, and protected application) during authentication and authorization in an OpenSSO Enterprise deployment [10].

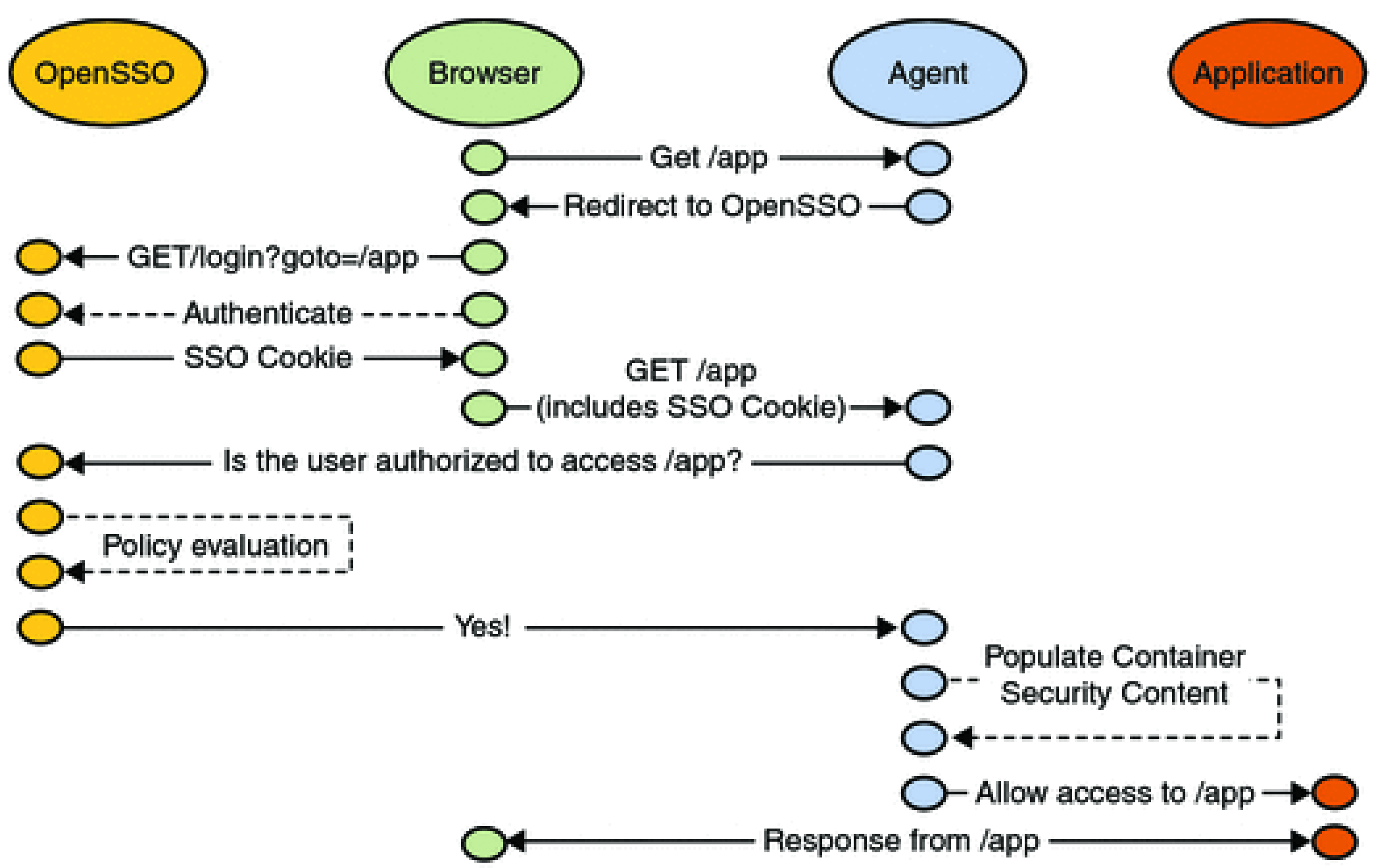

**Figure 2 : Open SSO Enterprise Authentication and Authorization Interactions**

**Conclusion**

As regulations push enterprises to rethink security strategy, and securing traffic in motion becomes a requirement, multiple encryption methods will be implemented to satisfy specific encryption standards. However, a new model is necessary to implement and manage a cohesive security strategy.
First and foremost, security policy must be consolidated to one entity. Today, security policy is split between all the technologies providing security services: firewalls; IDS/IPS; data protection; and, identity management. For data protection, common security policy should be in place to implement encryption whether application, SSL or IPSec. A common policy platform would enable a global set of rules such as resource entitlement (access based on groups of users, applications, or devices and implementation specifics).
Secondly, for data protection, key negotiation and exchange cannot limit network or application services. Encryption implementation today requires two end points to authenticate each other and exchange keying material. This sets up a point-to-point communications tunnel between the end points. As the need for data protection implementation grows the scalability of this approach is questionable. Imagine point-to-point tunnels to hundreds, if not thousands, of end points. Point-to-point key management is extremely difficult at best, and impossible in large mesh networks tying together thousands of end users.
The security model must separate key management from the end point devices. Key management should: leverage policy rules to enable grouping of end points, storing and archiving keys; generate and distribute keys to end points; and, provide the security policy interface to the end points.

Third, we need to start looking at security end points as any device or application (PDA, cell phone, software, router or switch). As we move to a security model where all end points are security enforcement points, the model needs to accommodate any type of device or software and reduce complexity as much as possible.
This security model would appear as:
The model leverages a common policy of separate encryption key management, thus improving data protection. New technologies and improved enterprise data protection architecture are necessary to provide this model of data protection.